\documentclass[11pt]{article}
\usepackage{geometry} 
\usepackage[latin1]{inputenc}               
\usepackage{graphicx}
\usepackage{bm}
\usepackage{amssymb}
\usepackage{epstopdf}
\usepackage{color}
\usepackage{etaremune} 
\usepackage{authblk}
\usepackage[T1]{fontenc}
\usepackage{breakcites}
\usepackage{multicol}
\usepackage[round]{natbib}
\graphicspath{{./figuras/}}

\date{}

\begin{document}
	 \title{\textbf{Comparing statistical methods to predict leptospirosis incidence using hydro-climatic covariables.}}
	\author[1]{Mar\'ia Jos\'e Llop}
	\author[2,3]{Pamela Llop}
	\author[1,3]{Mar\'ia Soledad L\'opez}
	\author[1,3]{Andrea G\'omez}
	\author[1,3]{Gabriela V. M\"uller}
	\affil[1]{CEVARCAM, Facultad de Ingenier\'ia y Ciencias H\'idricas, Universidad Nacional del Litoral (UNL), Santa Fe, Argentina}
	\affil[2]{Facultad de Ingenier\'ia Qu\'imica - Universidad Nacional del Litoral (UNL), Santa Fe, Argentina}
	\affil[3]{CONICET, Santa Fe, Argentina}

	\maketitle

\begin{abstract}
Leptospiroris, the infectious disease caused by the spirochete bacteria \textit{Leptospira interrogans}, constitutes an important public health problem all over the world. In Argentina, some regions present climate and geographic characteristics that favors the habitat of the bacteria \textit{Leptospira}, whose survival strongly depends on climatic factors. For this reason, regional public health systems should include, as a main factor, the incidence of the disease in order to improve the prediction of potential outbreaks, helping to stop or delay the virus transmission. The classic methods used to perform this kind of predictions are based in autoregressive time series tools which, as it is well known, perform poorly when the data do not meet their requirements. Recently, several  nonparametric methods have been introduced to deal with those problems. In this work, we compare a semiparametric method, called \textit{Semi-Functional Partial Linear Regression (SFPLR)} with the classic \textit{ARIMA} and a new alternative \textit{ARIMAX}, in order to select the best predictive tool for the incidence of leptospirosis in the Argentinian Litoral region. In particular, \textit{SFPLR} and \textit{ARIMAX} are methods that allow the use of (hydrometeorological) covariables which could improve the prediction of outbreaks of leptospirosis.
\end{abstract}


\section{Introduction}\label{intro}
Leptospirosis is a public health problem all over the world, particularly in tropical and subtropical areas. It is a zoonosis caused by the spirochete bacteria \textit{Leptospira interrogans} \citet{Cespedes05}. \textit{Leptospira} has been found in virtually all mammalian species examined. Humans most commonly become infected through occupational, recreational, or domestic contact with the urine of host animals, either directly or via contaminated water or soil (\citet{Adler10,Trueba04}). 
Infectious vector-borne diseases, particularly leptospirosis, are climatic-sensitive (\citet{WHOWMO12,Lopez18,Lopez19}). Therefore, extreme climate events enhanced by climate change increase the problems associated with people's health (\citet{Cerda08,Sanchez09}).
In this context, the average rainfall and extreme precipitation events both in intensity and frequency increased between 1951 and 2010 in northeastern Argentina (\citet{Barr09,Penalba10}). This region has important rivers such as Paran\'a and Uruguay, as a consequence, the highest rainfall caused significant flooding in the last decade (\citet{Antico14,Lopez18}) and this trend has continued to rise in recent years \citet{Lovino18b}. These extreme events influence the epidemiology of infectious diseases in the region (\citet{Lopez18,Lopez19}). For these reasons, health systems responses, besides the treatment of individual cases, should include the estimated number of cases of said disease. This estimation would improve the response of health systems during potential outbreaks, cutting of or delaying the virus transmission \citet{Canals10}.

The modeling of leptospirosis in relation to hydroclimatic variables has been studied using deterministicaly (\citet{Triampo07,Zaman10,Zaman12,Holt06}) and statistically (\citet{Chadsuthi12,Torres08}). For example, \citet{Chadsuthi12} modelled that rainfall is correlated with cases of leptospirosis in both of the studied regions, while temperature is correlated with the desease only in one of those regions. Models can be useful tools to show the trend of the variable of interest (in this case, the incidence of leptospirosis), by adjusting to the recorded data, or to predict the next seasonal peak with higher accuracy.

From the statistical point of view, the commonly used methods to predict cases of epidemiological diseases are the well known autorregresive models. The simplest of these models is the \textit{autorregresive (AR)} of order $p$ model, initially developed by \citet{Yule27}, which models linearly the response variable in terms of its own previous values and a stochastic term. The \textit{moving-average model (MA)}, introduced lately by \citet{Slutzky37}, proposes that the response variable depends linearly on the current and various past values of the stochastic term. The \textit{autoregressive moving-average (ARMA)} model (\citet{Wold38}), regress the response variable in terms of a linear combination of both, \textit{AR} and \textit{MA} models. More precisely, the \textit{AR} term models the response as a linear combination of its own past values and the \textit{MA} term involves modeling the error as a linear combination of previous values of the stochastic error. The \textit{autoregressive integrated moving average (ARIMA)} model (\citet{Box94}) is a generalization of the \textit{ARMA} to non-stationary series, meaning, those who mean and variance are non constant in time. The \textit{integrated} part refers to a differencing initial step, which can be applied in order to eliminate the non-stationarity of the serie. Some application of this method to epidemiological time series can be found, for instance, in \citet{Promprou06,Liu11,Countin07}. Although this model seems to be efficient, it only uses the variable of interest without taking into account the additional information that some predictor variables may present. As we mentioned before, when predicting outbreaks of leptospirosis in the Argentinian Litoral region, it is important to consider hydrometeorological covariables since they can be indicators of outbreaks and could improve the prediction. In this direction, the \textit{ARIMAX} model (see, for instance, \citet{Kongcharoen13}) is an extension of \textit{ARIMA} that proposes the use of covariables. Some applications of this method can be found in \citet{Chadsuthi12,Yogarajah13,avellaneda12}.

Although autoregressive predictive models are simple and easy to understand and analize, in real data applications they are  difficult to apply since, in this situations, rarely the data meets the (linear) requirements of this methods. For instance, the large amount of nulls values and the discrete nature of leptospirosis incidence data series, makes difficult to perform the analysis with the usual time series methods. In this direction, nonparametric and semiparametric methods are a preferable alternative to autoregressive models. Some references on semiparametric prediction of temporal series have been studied, among others, in \citet{Gyorfietal89}, \citet{Bosq97}, \citet{Bosq98} and \citet{Fanetal03}. Some reviews can also be found in \citet{Wolfgangetal08}. More recently, \citet{Biauetal10} studied the problem of sequential nonparametric prediction and proved some consistency results under mild conditions. In \citet{Shangetal11}, the authors introduced a nonparametric prediction method with dynamic updating and showed its performance using  monthly \textit{Ni\~no} indices from January, 1950 to December, 2008. 

In this paper we consider the method introduced by \citet{Aneirosetal08}, who proposed to treat time series readings for each year as a function, this way having, a sample of time series of functional data. More precisely, the idea behind this method is to split the observed long-time series in short-time trajectories, all of them of the same size (for instance, a year), resulting in a sample of curves, that can be then adapted to the functional data framework. 


That means, that although the data is recorded only in a discrete grid of time points, by adapting them to the functional data framework, we can assume a continuous underlying curve for each year.

The model by \citet{Aneirosetal08} is called \textit{Semi-Functional Partial Linear Regression (SFPLR)} and it is an additive model with two terms: one modeling \textit{nonparametrically} the (temporal) response variable and other adding the additional information presented in the covariates by \textit{linearly combining} them.  

As the authors mention, the \textit{SFPLR} method can have several advantages since, cutting the observed time series into a sample of curves and incorporating one single past trajectory rather than many single past values in the model, solves the problem of choosing the number of past values to be used in the construction of autoregressive prediction methods. In addition, in the context of infectious diseases with seasonal cycles depending on the climate, as it is the case of leptospirosis, this approach uses the measured value of the (hydroclimatic) covariables only in the month of interest which improves considerably the prediction. Another advantage of \textit{SFPLR} is that it is not necessary to predict a future value of the covariables since, to perform the prediction, the method uses only recorded past values of the variable of interest and covariables. 

In this work we compare de forecasting performance of the methods  \textit{ARIMA}, \textit{ARIMAX} and \textit{SFPLR}  when we applied them to historical leptospirosis data using, in addition, hydrometeorological information. The main goal in these kind of studies is select the best method 
 to predict the incidence of epidemiological diseases in the northeast of our country. In particular, we are interested in find the more suitable tool to predict outbreaks of leptospirosis that can be used by the regional public health systems.
 
The rest of the paper is organized as follows: in Section \ref{metodos} we describe the three prediction methods used in the applications, the source of data is presented in Section \ref{datos},  Section \ref{resultados} is devoted to presenting the results obtained when applying the prediction methods to leptospirosis data and finally, in Section \ref{conclusiones} we present the conclusions of the work. 

\section{Methods}\label{metodos}
In this Section we present the models that caracterize the three prediction methods: \textit{ARIMA}, \textit{ARIMAX} and \textit{SFPLR}, that we applied to the leptospirosis incidence data collected in the Litoral area of our country. All the numerical implementation was performed in the statistical software R \citet{R}.

\subsection{ARIMA model}
Following \citet{Box94}, we define the \textit{ARMA model} for stationary time series, as
\begin{equation}\label{Ec1}
 \Phi_p(B)Y_t = \Theta_q(B) a_t,
\end{equation}
where $B$ is the \textit{backward shift operator} which verifies $Y_{t-k}=B^kY_t$, $\Phi_p(B)$ is the \textit{autoregressive operator}, $\Theta_q(B)$ is the \textit{moving average operator} and  $a_t$ is a white noise process with $\mathbb{E}(a_t)=0$ and $\mathbb{V}(a_t)=\sigma^2_a$. 

Recall that a serie is \textit{(strictly) stationary} if, for any set of times $t_1,...,t_m$ and its lag $t_{1+d},...,t_{m+d}$ the joint distribution of the $m$ observations $z_{t_1},...,z_{t_m}$ and $z_{t_{1+d}},...,z_{t_{m+d}}$ is the same. For linear stochastic process, this is equivalent to say that roots of $\Phi_p(B)$ have a modulus (absolute value) less than one. 

In the case of non stationary processes, this is, when $\Phi_p(B)$ has unit roots, the process will need to be differenced multiple times ($d$) to become stationary and, in this case, the autoregressive operator can be written as $\Phi_p(B) = (1-B)^d \bm{\varphi}_p(B)$. In this case, in Equation (\ref{Ec1}) we get
\begin{equation}
\label{Ec2} \bm{\varphi}_p(B) (1-B)^d  Y_t = \Theta_q(B) a_t. 
\end{equation}
Observe that the model described before is the \textit{ARIMA(p,q,d) model} and corresponds to assume that, the difference of order $d$, $W_t=(1-B)^d  Y_t$ can be represented as the \textit{ARMA(p,q) model}
$$ 
\bm{\varphi}_p(B) W_t = \Theta_q(B) a_t. 
$$
The notation used for this kind of models is \textit{ARIMA(p,q,d)}, where $p$, $q$ and $d$ were described before and they indicate the number of autoregressive terms, the number of lagged forecast errors in the prediction equation and the number of nonseasonal differences needed for stationarity, respectively.


To adjust the ARIMA model, the following steps must be performed: 
\begin{itemize}
\item Verify the stationarity of the data using the known Dickey-Fuller test. If they are nonstationary, use Box-Cox transformation to stabilize the variance and make ($d$) differences between the observations to stabilize the mean.
\item Using autocorrelation and partial autocorrelation graphs, identify the orders $p$ and $q$ such that the serie $W_t$ can be modeled as an \textit{ARMA(p,q) model}. 
\end{itemize}

\subsection{ARIMAX model} 
As it was mentioned before, the \textit{ARIMAX model} is an extension of the \textit{ARIMA model} which allows the use of covariables. The expression that represent this model is
$$
\bm{\varphi}_p(B) (1-B)^d  Y_t = \theta(B)X_t + \Theta_q(B) a_t,
$$ 
where, analogous to the \textit{ARIMA} model, $B$ is the \textit{backward shift operator} which verifies $Y_{t-k}=B^kY_t$, $\bm{\varphi}_p(B)$ is the autoregressive stationary process, $X_t$ represents the covariables, $\Theta_q(B)$ is a tranference function representing the effect of $X_t$ on $Y_t$ and $\theta(B)$ is the moving average operator. 

\subsection{SFPLR model}

Following \citet{Aneirosetal08}, we assume that the long-time serie $Y_t$, has been observed at $N$ equispaced point times (this is, $t=1,\ldots,N$)  and we cut it (without losing generality) into $n$ short-times curves $Y_i$, $i=1,\dots,n$ of length $\tau$. As a consequence, we get a sample $\{(\bm{X}_i,Y_i)\}_{i=1}^n$ where for any curve $Y_i$, the vector $\bm{X}_i=(X_{i1},X_{i2},...,X_{ip})$ represents the numerical, independent, covariables. Let $G(Y_{i+1})$ some characteristic from the period of time $i+1$ that we want to study, for instance the leptospirosis incidence in an specific month. Then, the \textit{SFPLR model} is given by 
$$
G(Y_{i+1})= \sum_{j=1}^{p} X_{ij} \beta_j + m(Y_i) + \epsilon_i, \qquad  i=1,\dots,n,
$$ 
where $\boldsymbol{\beta} =(\beta_1, \beta_2,..., \beta_p)^T$ is a vector of knonwn real parameters, $m$  is an unknown (smooth) real function and $\epsilon_i$ are identically distributed random errors with $\mathbb{E}(\epsilon_i|X_{i1},X_{i2},...,X_{ip})=0$. The estimators (see \citet{kersmooth}) are given by 
$$
\widehat{\bm{\beta}}_h  = \left( \widetilde{\bm{X}}_h^T \widetilde{\bm{X}}_h\right) ^{-1} \widetilde{\bm{X}}_h^T \widetilde{\bm{Y}}_h,
$$
where $\widetilde{\bm{X}}_h= (\textbf{I}-\textbf{W}_h)\bm{X}$ and $\widetilde{\bm{Y}}_h= (\textbf{I}-\textbf{W}_h)\bm{Y}$, and 
$$
\hat{m}_h(t)=\sum_{i=1}^{n} w_{n,h}(t,Y_i)(Y_i-\bm{X}_i^T \widehat{\bm{\beta}}_h ),
$$
where $\bm{W}_h=(w_{n,h}(Y_i,Y_j))_{ij}$, $i,j=1,\ldots, n$ are Nadaraya-Watson type kernel weights with the smoothing parameter $h$ is chosen by cross-validation. 

\section{Data}\label{datos}
Leptospirosis incidence is Notified in Santa Fe and Entre R\'ios provinces from 2009, given that year \textit{Sistema Nacional de Vigilancia Epidemiol\'ogica por Laboratorios de Argentina} (SIVILA) was implemented. The confirmed cases of leptospirosis were requested to 
\textit{Direcci\'on de Promoci\'on y prevenci\'on de la Salud, Ministerio de Salud} of the province of Santa Fe and \textit{Divisi\'on Epidemiol\'ogica, Ministerio de Salud} 
of the province of Entre R\'ios. Therefore, the period of analysis is 2009-2018. The total number of confirmed leptospirosis cases was 810; 496 for Santa Fe and 314 for Entre R\'ios, for the mentioned period. 
\\
Selected covariables are those identified in \citet{Lopez19} as the main hydroclimatic indicators that can influence leptospirosis outbreaks occurrence in the northeastern Argentina. Hydroclimatic datasets identified in \citet{Lopez19} include monthly total precipitation, monthly river hydrometric levels and the Oceanic Ni\~no Index (ONI,NOAA/NWS/CPC). Precipitation data were provided by \textit{Servicio Meteorol\'ogico Nacional} (SMN)
and \textit{Instituto Nacional de Tecnolog\'ia Agropecuaria} (INTA).
The meteorological stations include: Sauce Viejo Aero, Rosario Aero and Paran\'a. 
Hydrometric data were provided by the 
\textit{Instituto Nacional del Agua} (INA) while hydrometric river evacuation level (hydrometric river level from which people are evacuated to safe non-floodable areas) data were provided by
\textit{Prefectura Naval Argentina} (PNA). ONI characterizes the years and months under El Ni\~no, La Ni\~na or neutral conditions. The ONI is the 3-month running mean SST anomaly for the Ni\~no 3.4 region (https://ggweather.com/enso/oni.htm).
Finally, data about population indicators were retrieved from the 2010 \textit{Instituto Nacional de Estad\'istica y Censos} (INDEC) Census.

\section{Empirical Results and Discussion}\label{resultados}

\begin{figure}[t!]
\begin{center}
\includegraphics[scale=0.35]{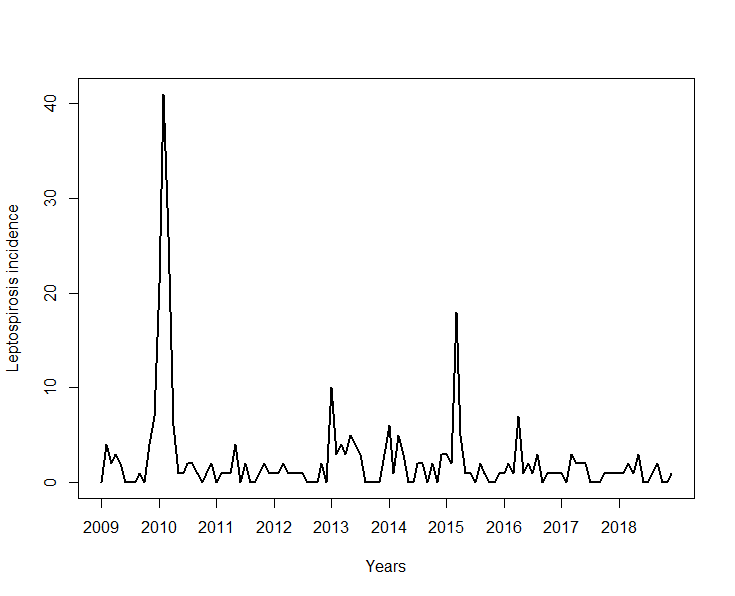}
\includegraphics[scale=0.35]{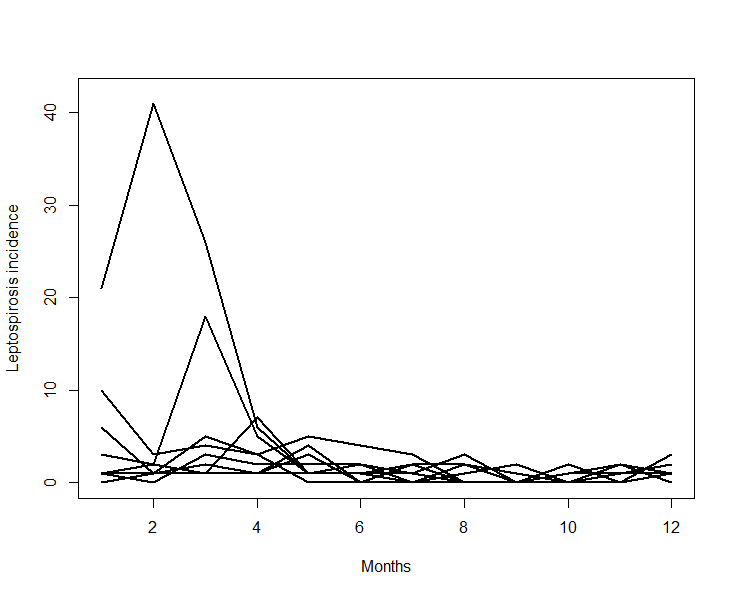}
\caption{Left: long-time serie of monthly leptospirosis incidence in the area of study. Right: short-time series of functional data.}\label{plot-lepto}
\end{center}
\end{figure}

\begin{figure}[t!]
	\begin{center}
		\includegraphics[scale=0.35]{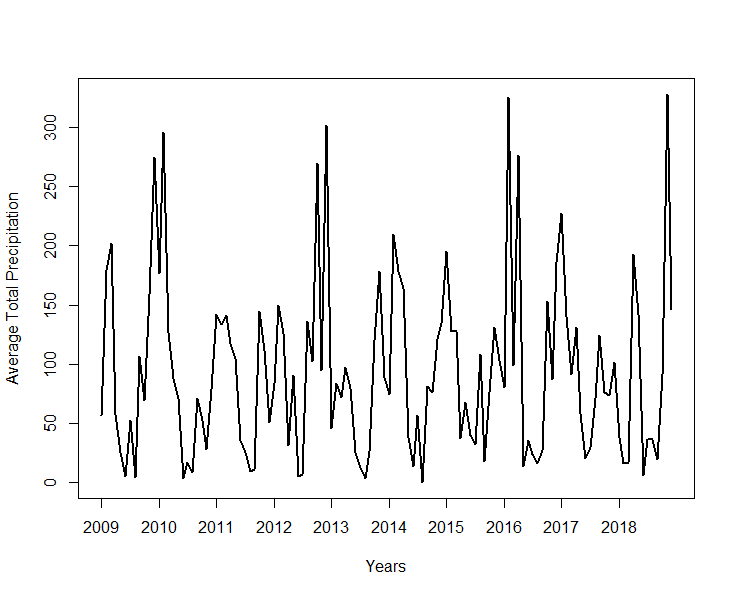}
		\includegraphics[scale=0.35]{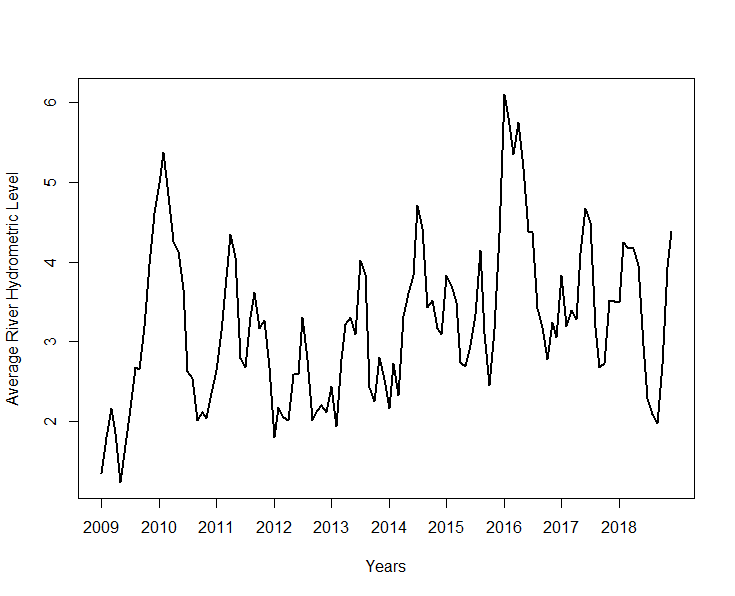}
		\includegraphics[scale=0.35]{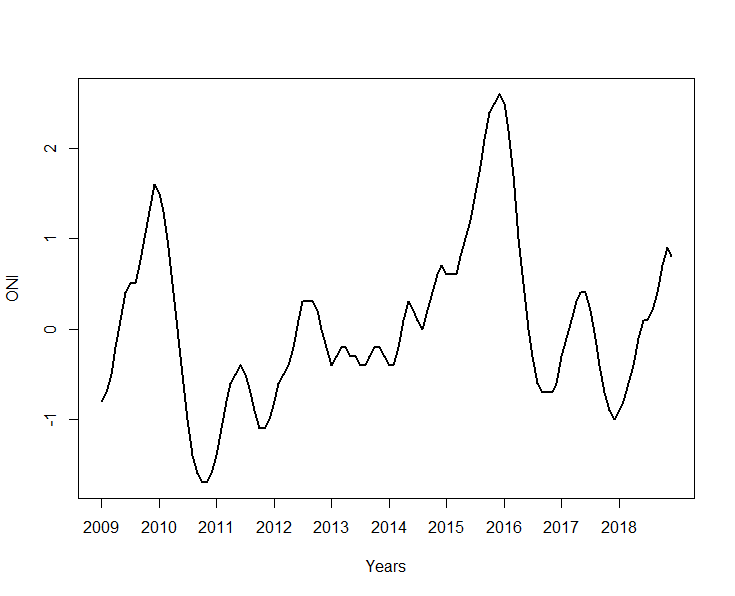}
		\caption{Hydroclimatic covariables}\label{variables}
	\end{center}
\end{figure}

To explore the observed leptospirosis data in the area of study, in Figure \ref{plot-lepto} we plot the long-time (left) and short-time (right) incidence series.
However, since in the area of study the climatic conditions can differ, we perform the statistical analysis for each region separately. Therefore, the statistical analysis was performed in the three selected cities.
To do that, for each region, we split the whole sample in two: the training and testing samples. With the former one, we estimate all the parameters involved in each model so that, we have the estimated predictive rules. This sample contains the leptospirosis incidence and the hydroclimatic covariables monthly total precipitation, monthly river hydrometric levels and the ONI Index from January, 2009 to December 2017. The time series of hydroclimatic covariables are plotted in Figura (\ref{variables}). 

The testing sample corresponds to the leptospirosis incidence from January, 2018 to December 2018 ($12$ months). With this sample we measure the predictive power the three methods: we evaluate the estimated rule in the testing sample in order to get the estimated leptospirosis incidence ($\hat{Y_i}$, $i=1,\dots,12$) and compare them with the true values ($Y_i$, $i=1,\dots,12$).  To compare the methods we used two criterion which are described below. 

The Nash--Sutcliffe criterion is
$$ 
\mbox{NSE}=1-\frac{\sum_{i=1}^{12}{(\hat{Y_i}-Y_i)^2}}{\sum_{i=1}^{12}{(\bar{Y}-Y_i)^2}},
$$
Observe that Nash--Sutcliffe criterion can range from $-\infty$ to $1$: $\mbox{NSE} = 1$ corresponds to a perfect prediction, $\mbox{NSE} = 0$ indicates that the method perform as well as the mean of the data, $\mbox{NSE} < 0$ occurs when the observed mean is a better predictor than the model. 

The second criterion is the Root Mean Square Error
$$ 
\mbox{RMSE}= \sqrt{\left( \frac{1}{12} \sum_{i=1}^{12} (\hat{Y_i}-Y_i)^2 \right)}.
$$

In Table \ref{tabla1} we present the results, best results are reported in bold.
\begin{table}
\begin{center}
\begin{tabular}{|c|c|c|}
\multicolumn{3}{c}{Santa Fe}\\
\hline
Method & NSE & $\mbox{RMSE}$\\ 
\hline
\hline
ARIMA  & -5.00 & 0.91 \\
\hline
ARIMAX  & -9.20   & 1.19\\
\hline
SFPLR  & \textbf{-1.40} & \textbf{0.58} \\
\hline
\end{tabular}
\hspace{0.5cm}
\begin{tabular}{|c|c|c|}
\multicolumn{3}{c}{Paran\'a}\\
\hline
Method & NSE & $\mbox{RMSE}$\\ 
\hline
\hline
ARIMA  & -2.38 &  0.87 \\
\hline
ARIMAX  & -1.63  & 0.76 \\
\hline
SFPLR  & \textbf{-0.50} & \textbf{0.58} \\
\hline
\end{tabular}
\begin{tabular}{|c|c|c|}
\multicolumn{3}{c}{Rosario}\\
\hline
Method & NSE & $\mbox{RMSE}$ \\ 
\hline
\hline
ARIMA  & \textbf{-0.63} & \textbf{0.82} \\
\hline
ARIMAX  & -0.83   & 0.87 \\
\hline
SFPLR  & \textbf{-0.63}  & \textbf{0.82} \\
\hline
\end{tabular}
\end{center}
\caption{NSE and MSE for each region.} \label{tabla1}
\end{table}

\section{Conclusions}\label{conclusiones}

In this work we compare de forecasting performance of the methods \textit{ARIMA}, \textit{ARIMAX} and \textit{SFPLR}, when we applied them to leptospirosis data collected in the different regions, in addition, hydrometeorological information. We found that, in general, \textit{SFPLR} provides better forecasting performance than \textit{ARIMA} and \textit{ARIMAX}, except in the region of Gualeguach\'u, were there was no cases of leptospirosis, this is, the data is almost al s zero. 

It is worth to say that, although for this kind of data (incidence) the good performance of \textit{SFPLR} is not very visible, in datasets that are continuous and not so sparse (this is, with less zeros) like the Ozone dataset presented in \citet{Aneirosetal08}, compared with \textit{ARIMA} and \textit{ARIMAX}, \textit{SFPLR} perform highly better (see Table \ref{tabla2}).

In Table \ref{tabla1} are showed the values of both criterion obtained for 2018 in each of the three cities, for the ARIMA, ARIMAX and SFPLR models implementations.

In general, SFPLR provides better forecasting performance than ARIMA and ARIMAX, except in the Rosario case where similar values are obtained for SFPLR and ARIMA.

In general, SFPLR provides better forecasting performance than ARIMA and ARIMAX in the three studied cities.

SFPLR is able to predict the future leptospirosis occurrence relatively accurate considering hydroclimatic variables. However, leptospirosis is also determined by social factors, such as human activities, age and sex of individuals and health system prophylaxis that might increase the probability of infection (citas). Some of these factors could be considered in future studies.

Through this first comparative analysis, could be concluded that these types of models become a set of suitable tools to predict outbreaks of leptospirosis in northeastern Argentina and could
be used by the regional public health systems.

%

\begin{table}
\begin{center}
\begin{tabular}{|c|c|c|}
\hline
Method & NSE & $\sqrt{\mbox{MSE}}$\\ 
\hline
\hline
ARIMA  & -0.05 & 36.65 \\
\hline
ARIMAX  & 0.85   & 13.73\\
\hline
SFPLR  & \textbf{0.93} & \textbf{9.70} \\
\hline
\end{tabular}
\end{center}
\caption{NSE and MSE for the Ozone data set.} \label{tabla2}
\end{table}

\bibliographystyle{plainnat}
\bibliography{Bibs}

\begin{thebibliography}{39}
\providecommand{\natexlab}[1]{#1}
\providecommand{\url}[1]{\texttt{#1}}
\expandafter\ifx\csname urlstyle\endcsname\relax
  \providecommand{\doi}[1]{doi: #1}\else
  \providecommand{\doi}{doi: \begingroup \urlstyle{rm}\Url}\fi

\bibitem[Adler and Moctezuma(2010)]{Adler10}
B.~Adler and A.~Moctezuma.
\newblock Leptospira and leptospirosis.
\newblock \emph{Veterinary Microbiology}, 140:\penalty0 287--296, 2010.

\bibitem[Aneiros-P\'erez and Vieu(2008)]{Aneirosetal08}
G.~Aneiros-P\'erez and P.~Vieu.
\newblock Nonparametric time series prediction: a semi-functional partial
  linear modelling.
\newblock \emph{Journal of Multivariate Analysis}, 99\penalty0 (5):\penalty0
  834--857, 2008.
\newblock URL \url{https://doi.org/10.1016/j.jmva.2007.04.010}.

\bibitem[Antico et~al.(2014)Antico, Schlotthauer, and Torres]{Antico14}
A.~Antico, G.~Schlotthauer, and M.E. Torres.
\newblock Analysis of hydroclimatic variability and trends using a novel
  empirical mode decomposition: Application to the paran\'a river basin.
\newblock \emph{Journal of Geophysical Research Atmospheres}, 119:\penalty0
  1218--1233, 2014.

\bibitem[Avellaneda et~al.(2012)Avellaneda, Ochoa, and
  Figueroa~Garc\'ia]{avellaneda12}
J.A. Avellaneda, C.M. Ochoa, and J.C. Figueroa~Garc\'ia.
\newblock Comparaci\'on entre un sistema neuro difuso auto organizado y un
  modelo arimax en la predicci\'on de series econ\'omicas vol\'atiles.
\newblock \emph{Ingenier\'ia}, 17\penalty0 (2):\penalty0 26--34, 2012.

\bibitem[Barros(2009)]{Barr09}
V.R. Barros.
\newblock Extreme rainfalls in se south america.
\newblock \emph{Climatic Change}, 1-2\penalty0 (96):\penalty0 119--136, 2009.

\bibitem[Biau et~al.(2010)Biau, Bleakley, Gy\"orfi, and Ottucs\'ak]{Biauetal10}
G.~Biau, K.~Bleakley, L.~Gy\"orfi, and G.~Ottucs\'ak.
\newblock Nonparametric sequential prediction of time series.
\newblock \emph{Journal of Nonparametric Statistics}, 22:\penalty0 297--317,
  2010.

\bibitem[Bosq(1997)]{Bosq97}
D.~Bosq.
\newblock Parametric rates of nonparametric estimators and predictors for
  continuous time processes.
\newblock \emph{The Annals of Statistics}, 25:\penalty0 982--1000, 1997.

\bibitem[Bosq(1998)]{Bosq98}
D.~Bosq.
\newblock \emph{Nonparametric statistics for stochastic processes. Estimation
  and Prediction}.
\newblock Lecture Notes in Stat., Vol. 110, 2nd. ed., Springer, New York, 1998.

\bibitem[Box et~al.(1994)Box, Jenkins, and Reinsel]{Box94}
G.~Box, G.~Jenkins, and G.~Reinsel.
\newblock \emph{Time Series Analysis: Forecasting and Control. 3th ed.}
\newblock Prentiuce Hall Canada, 1994.

\bibitem[C. et~al.(2008)C., Lele, Pascual, Bouma, and IKo]{Torres08}
Torres C., S~Lele, M.~Pascual, M~Bouma, and A.~IKo.
\newblock A stochastic model for ecological systems with strong nonlinear
  response to environmental drivers: application to two water-borne diseases.
\newblock \emph{Journal of the Royal Society Interface}, 5:\penalty0 247--252,
  2008.

\bibitem[Canals(2010)]{Canals10}
M.L. Canals.
\newblock Predictibilidad a corto plazo del n\'umero de casos de la influenza
  pand\'imica ah1n1 basada en modelos determin\'isticos.
\newblock \emph{Revista Chilena de Infectolog\'ia}, 27\penalty0 (2):\penalty0
  119--125, 2010.

\bibitem[Cerda et~al.(2008)Cerda, Valdivia, Valenzuela, and Venegas]{Cerda08}
L.J. Cerda, C.G. Valdivia, B.M. Valenzuela, and L.J. Venegas.
\newblock Cambio clim\'atico y enfermedades infecciosas. un nuevo escenario
  epidemiol\'ogico.
\newblock \emph{Revista Chilena de Infectolog\'ia}, 25\penalty0 (6):\penalty0
  447--452, 2008.

\bibitem[C\'espedes(2005)]{Cespedes05}
M.Z. C\'espedes.
\newblock Leptospirosis: Enfermedad zoon\'otica emergente.
\newblock \emph{Revista Peruana de Medicina Experimental y Salud Publica},
  22\penalty0 (4):\penalty0 290--307, 2005.

\bibitem[Chadsuthi et~al.(2012)Chadsuthi, Modchang, Lenbury, Iamsirithaworn,
  and Triampo]{Chadsuthi12}
S.~Chadsuthi, C.~Modchang, Y.~Lenbury, S.~Iamsirithaworn, and W.~Triampo.
\newblock Modeling seasonal leptospirosis transmission and its association with
  rainfall and temperature in thailand using time-series and arimax analyses.
\newblock \emph{Asian Pacific Journal of Tropical Medicine}, 5:\penalty0
  539--546, 2012.

\bibitem[Cout\'in(2007)]{Countin07}
M.G. Cout\'in.
\newblock Utilizaci\'on de modelos arima para la vigilancia de enfermedades
  transmisibles en cuba.
\newblock \emph{Revista Cubana Salud P\'ublica}, 33\penalty0 (1), 2007.

\bibitem[Eugen(1937)]{Slutzky37}
S.~Eugen.
\newblock The summation of random causes as the source of cyclic processes.
\newblock \emph{Econometrica}, 5\penalty0 (2):\penalty0 105--146, 1937.

\bibitem[Fan and Yao(2003)]{Fanetal03}
J.~Fan and Q.~Yao.
\newblock \emph{Nonlinear Time Series. Nonparametric and Parametric Methods}.
\newblock Springer Series in Statistics, Springer, New York, 2003.

\bibitem[Gy\"orfi et~al.(1989)Gy\"orfi, H\"ardle, Sarda, and
  Vieu]{Gyorfietal89}
L.~Gy\"orfi, W.~H\"ardle, P.~Sarda, and P.~Vieu.
\newblock \emph{Nonparametric curve estimation from time series}.
\newblock Springer-Verlag, Berlin, 1989.

\bibitem[H. et~al.(1997)H., L., and C.]{Wolfgangetal08}
Wolfgang H., Helmut L., and Rong C.
\newblock A review of nonparametric time series analysis.
\newblock \emph{International Statistical Review}, 65:\penalty0 49--72, 1997.

\bibitem[Holt et~al.(2006)Holt, Davis, and Leirs]{Holt06}
J.~Holt, S.~Davis, and H.~Leirs.
\newblock A model of leptospirosis infection in an african rodent to determine
  risk to humans: Seasonal fluctuations and the impact of rodent control.
\newblock \emph{Acta Tropica}, 99\penalty0 (2--3):\penalty0 218--225, 2006.

\bibitem[Kongcharoen and Kruangpradit(2013)]{Kongcharoen13}
Chaleampong Kongcharoen and Tapanee Kruangpradit.
\newblock Autoregressive integrated moving average with explanatory variable
  (arimax) model for thailand export.
\newblock 2013.

\bibitem[Liu et~al.(2011)Liu, Liu, Jiang, and Yang]{Liu11}
Q.~Liu, X.~Liu, B.~Jiang, and W.~Yang.
\newblock Forecasting incidence of hemorrhagic fever with renal syndrome in
  china using arima model.
\newblock \emph{BMC infectious diseases}, 11\penalty0 (1):\penalty0 218, 2011.

\bibitem[L\'opez et~al.(2018)L\'opez, M\"uller, and Sione]{Lopez18}
M.S. L\'opez, G.~M\"uller, and W.~Sione.
\newblock Analysis of the spatial distribution of scientific publications
  regarding vector-borne diseases related to climate variability in south
  america.
\newblock \emph{Spatial and Spatio-temporal Epidemiology}, 26:\penalty0 35--93,
  2018.

\bibitem[L\'opez et~al.(2019)L\'opez, M\"uller, Lovino, G\'omez, Sione, and
  Aragon\'es~Pomares]{Lopez19}
M.S. L\'opez, G.V. M\"uller, M.A. Lovino, A.A. G\'omez, E.F. Sione, and
  L.~Aragon\'es~Pomares.
\newblock Spatio-temporal analysis of leptospirosis incidence and its
  relationship with hydroclimatic indicators in northeastern argentina.
\newblock \emph{Science of the Total Environment}, 694, 2019.

\bibitem[Lovino et~al.(2018)Lovino, M\"uller, Berbery, and M\"uller]{Lovino18b}
M.A. Lovino, O.~M\"uller, E.H. Berbery, and G.~M\"uller.
\newblock How have daily climate extremes changed in the recent past over
  northeastern argentina?
\newblock \emph{Global and Planetary Change}, 168:\penalty0 78--97, 2018.

\bibitem[Organization and Organization(2012)]{WHOWMO12}
World~Health Organization and World~Meteorological Organization.
\newblock Atlas of health and climate.
\newblock \emph{World Health Organization}, 2012.

\bibitem[Penalba and Robledo(2010)]{Penalba10}
O.~Penalba and F.~Robledo.
\newblock Spatial and temporal variability of the frequency of extreme daily
  rainfall regime in the la plata basin during the 20th century.
\newblock \emph{Climatic Change}, 98:\penalty0 531--550, 2010.

\bibitem[Promprou et~al.(2006)Promprou, Jaroensutasinee, and
  Jaroensutasinee]{Promprou06}
S.~Promprou, M.~Jaroensutasinee, and K.~Jaroensutasinee.
\newblock Forecasting dengue haemorrhagic fever cases in southern thailand
  using arima models.
\newblock \emph{Dengue Bulletin}, 2006.

\bibitem[{R Core Team}(2013)]{R}
{R Core Team}.
\newblock \emph{R: A Language and Environment for Statistical Computing}.
\newblock R Foundation for Statistical Computing, Vienna, Austria, 2013.
\newblock URL \url{http://www.R-project.org/}.

\bibitem[S\'anchez et~al.(2009)S\'anchez, Liliana, Salim, and
  Gonz\'alez]{Sanchez09}
L.~S\'anchez, Liliana, M.~Salim, and M.~Gonz\'alez.
\newblock Cambios clim\'aticos y enfermedades infecciosas: Nuevos retos
  epidemiol\'ogicos.
\newblock \emph{Revista MVZ C\'ordoba}, 14\penalty0 (3):\penalty0 1876--1885,
  2009.

\bibitem[Shang and Hyndman(2011)]{Shangetal11}
H.L. Shang and R.J. Hyndman.
\newblock Nonparametric time series forecasting with dynamic updating.
\newblock \emph{Mathematics and Computers in Simulation}, 81\penalty0
  (7):\penalty0 1310--1324, 2011.

\bibitem[Speckman(1988)]{kersmooth}
P.~Speckman.
\newblock Kernel smoothing in partial linear models.
\newblock \emph{Journal of the Royal Statistical Society}, 50\penalty0
  (3):\penalty0 413--436, 1988.

\bibitem[Triampo et~al.(2007)Triampo, Baowan, Tang, Nuttavut, Wong-Ekkabut, and
  Doungchawee]{Triampo07}
W.~Triampo, D.~Baowan, I.M. Tang, N.~Nuttavut, J.~Wong-Ekkabut, and
  G.~Doungchawee.
\newblock A simple deterministic model for the spread of leptospirosis in
  thailand.
\newblock \emph{International Journal of Biological and Medical Sciences},
  2:\penalty0 22--26, 2007.

\bibitem[Trueba et~al.(2004)Trueba, Zapata, Madrid, Cullen, and
  Haake]{Trueba04}
G.~Trueba, S.~Zapata, K.~Madrid, P.~Cullen, and D.~Haake.
\newblock Cell aggregation: a mechanism of pathogenic leptospira to survive in
  fresh water.
\newblock \emph{International Microbiology}, 7\penalty0 (1):\penalty0 35--40,
  2004.

\bibitem[Wold(1938)]{Wold38}
H.~Wold.
\newblock \emph{A Study In The Analysis Of Stationary Time Series}.
\newblock Almqvist and Wiksells Boktryckert Uppsala, 1938.

\bibitem[Yogarajah et~al.(2013)Yogarajah, Elankumaran, and
  Vigneswaran]{Yogarajah13}
B~Yogarajah, C~Elankumaran, and R~Vigneswaran.
\newblock Application of arimax model for forecasting paddy production in
  trincomalee district in sri lanka.
\newblock 2013.

\bibitem[Yule(1927)]{Yule27}
G.U. Yule.
\newblock On a method of investigating periodicities in disturbed series, with
  special reference to wolfer's sunspot numbers.
\newblock \emph{Philosophical Transactions of the Royal Society of London.
  Series A, Containing Papers of a Mathematical or Physical Character},
  226:\penalty0 267--298, 1927.

\bibitem[Zaman(2010)]{Zaman10}
G.~Zaman.
\newblock Dynamical behavior of leptospirosis disease and role of optimal
  control theory.
\newblock \emph{International Journal of Mathematics and Computation},
  10\penalty0 (7):\penalty0 80--92, 2010.

\bibitem[Zaman et~al.(2012)Zaman, Khan, Islam, Chohan, and Jung]{Zaman12}
G.~Zaman, M.~Khan, S.~Islam, M.~Chohan, and I.~Jung.
\newblock Modeling dynamical interactions between leptospirosis infected vector
  and human population.
\newblock \emph{Applied Mathematical Sciences}, 6\penalty0 (26):\penalty0
  1287--1302, 2012.

\end{thebibliography}

\end{document}